\documentclass[aps,reprint,showpacs,floatfix,pre,twocolumn]{revtex4-1}
\usepackage{amsmath}
\usepackage{graphicx}
\usepackage{amssymb}
\usepackage{epsf}
\usepackage{color}
\usepackage{fancyhdr}
\usepackage{rotating}
\usepackage{longtable}
\usepackage{supertabular}
\usepackage{latexsym}
\usepackage{float}
\usepackage{dcolumn}
\usepackage{bm}

\newcommand\be {\begin{equation}}
\newcommand\ee {\end{equation}}
\newcommand\bea {\begin{eqnarray}}
\newcommand\eea {\end{eqnarray}}

\newcommand\bc {\begin{center}}
\newcommand\ec {\end{center}}
\newcommand\bfl{\begin{flushleft}}
\newcommand\efl{\end{flushleft}}
\newcommand\bfr{\begin{flushright}}
\newcommand\efr{\end{flushright}}
\def\bra{\langle}
\def\ket{\rangle}

\begin{document}

\title {Extending multiple histogram reweighting to a continuous lattice spin system
exhibiting a first order phase transition}
\author{Suman Sinha}%
\email{suman.sinha.phys@gmail.com}
\affiliation{Department of Physics, University of Calcutta, 
92 Acharya Prafulla Chandra Road, Kolkata 700009, India}

\begin{abstract}
We present extensive Monte Carlo simulations on a two-dimensional XY model 
with a modified form of interaction potential. Thermodynamic quantities other
than energy, specific heat etc (such as magnetization, susceptibility, fourth
order cumulant of magnetization) are obtained using multiple-histogram 
reweighting of the data obtained from the simulations. We employ an approach which 
eliminates the need to construct two-dimensional histograms. This approach makes 
judicious use of computer memory as well as CPU time. Lee-kosterlitz's method of 
finite size scaling for a first order transition and analysis using Binder's cumulant 
method allow us to make an accurate determination of the transition temperature.
\end{abstract}


\maketitle 

In $1988$, Ferrenberg and Swendsen showed that histograms can be used to extract
the maximum information from Monte Carlo (MC) data at a single temperature in 
the neighbourhood of a critical point \cite{fs1}. However, for studying phase
transitions, it is often desired to investigate the behaviours of a system over
a wide range of temperature values. In this situation, it is necessary to perform
simulations at more than one value of the temperatures of interest. In $1989$, 
Ferrenberg and Swendsen presented an optimized method for combining the data 
from an arbitrary number of simulations to obtain information over a wide range of
temperature values in the form of continuous functions \cite{fs2}. The method, known
as multiple histogram reweighting (MHR) method, provides a clear guide to 
optimize the length and location of additional simulations to provide maximum 
accuracy. The MHR method has the merit that it can be used with any simulation 
method that provides data for a system in equilibrium and it requires a negligible
amount of additional computer time for its implementation.

The MHR method allows us to interpolate results between several different simulations
performed at different temperatures. Suppose we want to estimate the average energy
$\bra E \ket$ over a range of temperatures. In the case of several simulations, the
upper end of one simulation's range is the lower end of another's. It should be 
possible to combine the estimates from the two simulations to give a better estimate
of $\bra E \ket$. Indeed, since every simulation gives an estimate (however poor) of
$\bra E \ket$ at every temperature, we should be able to combine all of these estimates
in some fashion (giving greater weight to those which are more accurate) to give the best
possible estimate for $\bra E \ket$, given our several simulations. This, in essence, is
the idea behind the MHR method. In this method, the data contained in the histograms of
energy are combined to yield an optimized estimate of the density of states
$\rho (E)$ and once $\rho (E)$ is known, $\bra E \ket$ can be estimated easily. The MHR 
method can be extended to provide interpolation of quantities other than average energy,
 for example average magnetization (order parameter) $\bra M \ket$, but this involves 
constructing two dimensional histograms. In this approach the data contained in histograms
of the energy and the magnetization from the simulations performed at different values of
temperatures are combined to yield an optimized estimate for the joint density of states $W (E, M)$.
The probability distribution $P (E, M)$ for an inverse temperature $\beta$ (where $\beta = 1/K_BT$,
$K_B$ being the Boltzmann constant set to unity) is then given by
\be
P (E, M)=\frac{1}{Z}W (E, M){\tt exp}(-\beta E)
\label{nprob}
\ee
where
\be
Z=\sum_{E, M}W (E, M){\tt exp}(-\beta E)
\label{pfn}
\ee
The estimate of the optimized density of states after Ref. \cite{fs2} obtained from $R$ simulations
performed at $\beta$ values $\beta_1$, $\beta_2$, $\cdots$ $\beta_R$ is given by
\be
W (E, M)=\frac{\sum_{i=1}^R g_i^{-1} N_i (E, M)}
{\sum_{i=1}^R g_i^{-1} n_i {\tt exp} [\beta_i (f_i-E)]}
\label{opdos}
\ee
where $g_i^{-1}$ is related to the auto correlation time $\tau_i$ of the $i^{th}$ simulation by
$g_i=1+2\tau_i$, $N_i (E, M)$ is the histogram count for the $i^{th}$ simulation, $n_i$ is the
length (in MCS) of simulation $i$ and $f_i$ is an estimate of free energy at $\beta = \beta_i$
and is determined self-consistently by iterating the relation
\be
{\tt exp}(-\beta_i f_i)=Z(\beta_i)=\sum_{E, M}W (E, M){\tt exp}(-\beta_i E)
\label{feng}
\ee
with $W (E, M)$ given by Eq. (\ref{opdos}). One MCS is taken to be completed when
the number of attempted single spin moves equals the number of spins in the system.
A good discussion of the MHR method may be found in \cite{nb}.

In practice, constructing two-dimensional histograms takes up a lot of computer memory
as well as being inappropriate for systems with continuous energy spectra. In continuous
lattice spin systems, one needs to use a discretization scheme to divide the energy range
of interest into a number of bins and because of the large number of bins
involved, it is inconvenient to work with the complete two-dimensional probability
distribution $P (E, M)$. To get rid of this difficulty, we have adopted a method \cite{flan}
which uses only the one-dimensional histograms $N_i (E)$ and have estimated for each energy (bin) the constant 
energy average of any function of $M$, $f(M)$ which we wish to study. In the present work,
we have evaluated the first, second and the fourth moments of magnetization distribution
which allows us to determine the average magnetization, susceptibility and Binder's fourth-
order cumulant of the system under investigation.

For the purpose of investigation, we have considered an extension of the two-dimensional
($2D$) XY model with a modified form of interaction potential introduced by Domany {\it et. al.}
\cite{domany}. The model consists of classical spins (of unit length), located at the sites of
a square lattice and are free to rotate in a plane, say the $X-Y$ plane (having no $Z$ component),
which interact with the nearest neighbours through a modified potential
\be
V(\theta)=2J\left[1-\left({\tt cos}^2 \frac{\theta}{2}\right)^{p^2}\right]
\label{pot}
\ee
where $\theta$ is the angle between the nearest neighbour spins, $J$ is the coupling constant
(conventionally set to unity)
and $p^2$ controls the non linearity of the potential well, although variation in $p^2$ does 
not disturb the essential symmetry of the Hamiltonian. For $p^2=1$, the potential reproduces
the conventional XY model which is known to exhibit a continuous transition of infinite order,
mediated by the unbinding of topological defects. This is the well-known Kosterlitz-Thouless
(KT) transition \cite{kt1,kt2}. For larger values of $p^2$ (say $p^2=50$), the model behaves
like a dense defect system \cite{ssskr3} and gives rise to a first order phase transition as
all the finite size scaling rules for a first order transition were seen to be obeyed \cite{ssskr2}.
It is to be mentioned in this context that van Enter and Shlosman provided a rigorous proof
\cite{enter1,enter2} of a first order phase transition in various SO($n$)-invariant $n$-vector models
that have a deep and narrow potential well. The model defined by Eq. (\ref{pot}) is a member of
these general class of systems.

The purpose of the present work is to test the efficiency and powerfulness
of the decades-old MHR method and its extension to determine thermodynamic quantities  
other than energy, specific heat etc by avoiding construction of $2D$ histograms, even
when applied to a lattice spin model with continuous energy spectra, exhibiting  a
sharp first order transition. The model considered is hard to simulate due to the 
occurence of a deep and narrow potential well for large values of $p^2$. We have applied this approach
to calculate magnetization, susceptibility and Binder's fourth order cumulant, quantities
that have not been estimated earlier for this model.
Our results and
analysis of data confirm the first order nature of transition by verifying the Lee and Kosterlitz's
method \cite{leekos} of finite size scaling for a first order phase transition. The transition 
temperature obtained from this study is also in agreement with previous studies.

Now we present results of the extensive MC simulations. 
We have used the single spin flip Metropolis algorithm \cite{metro} with some 
modifications in spin update scheme to obtain the raw data. 
The modifications have been discussed in Ref. \cite{ssskr3}.
We analyze our data from the Lee-Kosterlitz
method of finite size scaling \cite{leekos} and Binder's cumulant method \cite{bin,lanbin,plan,ts,lpb,clb,binheer} 
with optimized reweighting of data from multiple simulations to temperatures other than those at which
the simulations were performed. As a consequence of this approach we can accurately obtain the
transition temperature from the Binder's cumulant and determine the location and value of the maxima 
of susceptibility. The size of the energy bins is taken to be 0.004. 
We have checked that within statistical errors, the size of the bin did not
affect the numerical results of our simulation. In the simulations, $10^7$ MC steps per site were used
to compute the raw histograms and $10^6$ MC steps per site were taken for equilibration. The value of
$p^2$ is taken to be $50$ in this work.

\begin{figure}[!h]
\centering
\resizebox{80mm}{!}{\includegraphics[scale=0.6]{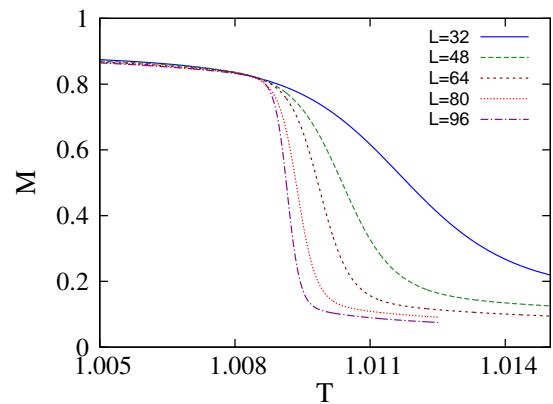}}
\caption{(Color online) The average magnetization $M$ plotted against dimensionless temperature
$T$ for different lattice sizes.}
\label{magfig}
\end{figure}

\begin{figure}[!h]
\centering
\resizebox{80mm}{!}{\includegraphics[scale=0.6]{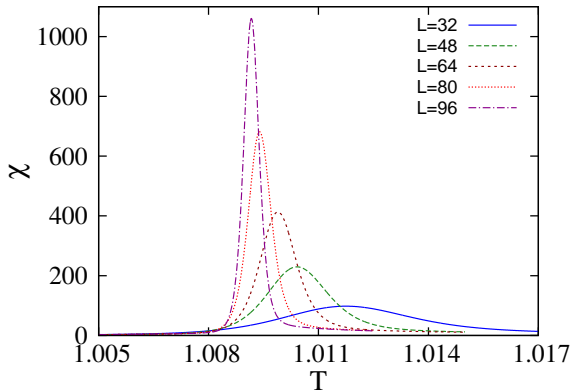}}
\caption{(Color online) The susceptibility $\chi$ plotted against dimensionless temperature
$T$ for different lattice sizes.}
\label{susfig}
\end{figure}

\begin{figure}[!h]
\centering
\resizebox{80mm}{!}{\includegraphics[scale=0.6,angle=-90]{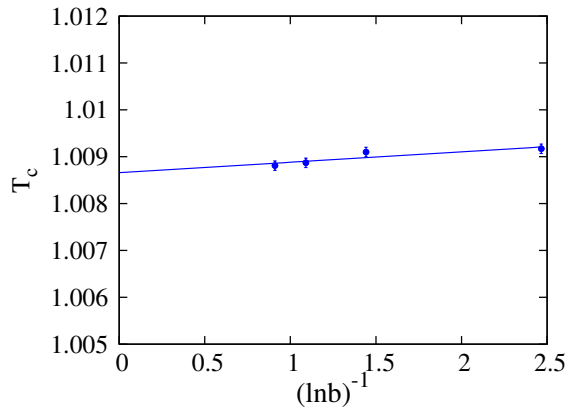}}
\caption{(Color online) Estimates of $T_c$ plotted against inverse logarithm of the scale
factor $b=L^{\prime}/L$. The solid line gives the best linear fit to the data points. 
The intercept on the $Y$ axis is $1.00866 \pm 0.00013$.}
\label{bincutc}
\end{figure}

Fig. \ref{magfig} shows the temperature variation in the magnetization ($M$) for a number of lattices, as
is obtained by applying MHR method described earlier.
It is evident from Fig. \ref{magfig} that with the increase in lattice size, the drop in the 
magnetization becomes sharper with the increase in temperature. The susceptibility $\chi$, which is
fluctuations in magnetization, as a function of temperature for various lattice sizes are displayed
in Fig. \ref{susfig}.
The transition is manifested by a huge peak height in $\chi$ and the data display a divergent
behaviour with increasing $L$, which is indicative of a discontinuous jump in $M$ in an 
infinite lattice. The finite size scaling of $\chi$ is now presented.
\begin{figure}[!h]
\centering
\resizebox{80mm}{!}{\includegraphics[scale=0.6,angle=-90]{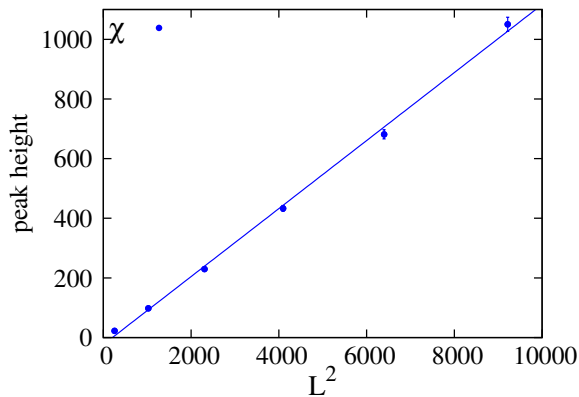}}
\caption{(Color online) Peak heights of $\chi$ plotted against $L^2$ with the linear
fit represented by the straight line. The error bars for most points are smaller than
the dimension of the symbols used for plotting.}
\label{fsssus}
\end{figure}
From Fig. \ref{fsssus}, where the maxima of $\chi$ are plotted against $L^2$, it is clear
that the standard scaling rules $\chi \sim L^d$ for a first order transition \cite{leekos}
are accurately obeyed in this model. We have also tested the finite size scaling relation
\be
T_c(L)-T_c(\infty) \sim L^{-d}
\label{fsstempeq}
\ee
which is valid for a first order phase transition \cite{leekos}. $T_c(\infty)$ represents the
thermodynamic limit of the transition temperature $T_c$ and $d$ is the spatial dimensionality
of the system. The transition temperature is estimated from the peak position of the 
susceptibility $\chi$.
\begin{figure}[!h]
\centering
\resizebox{80mm}{!}{\includegraphics[scale=0.6,angle=-90]{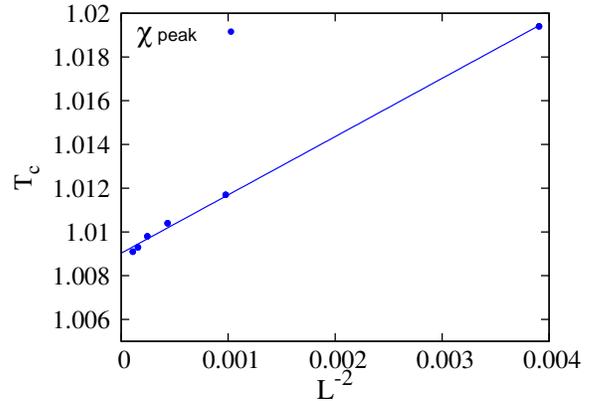}}
\caption{(Color online) The transition temperature $T_c$ obtained from susceptibility peak
position plotted against $L^{-2}$ along with the linear fits. The intercept on the $Y$ axis is
$1.00903 \pm 9 \times 10^{-5}$. The error bars are of the dimension of the symbols used for
plotting.}
\label{fsstemp}
\end{figure}
In Fig. \ref{fsstemp} the transition temperatures thus obtained have been plotted against
$L^{-2}$. It is seen that the linear fit is good within statistical errors and the thermodynamic
limit of the transition temperature is $1.00903 \pm 9 \times 10^{-5}$. We now focus our attention
on the study of the behaviour of the Binder's cumulant. Properties of the fourth-order
cumulants of magnetization are quite effective in characterizing phase transitions
\cite{bin,lanbin,plan,ts,lpb,clb,binheer}. 
 It is defined by
\be
V_L=1-\frac{\bra M^4 \ket_L}{3\bra M^2 \ket_L^2}
\label{bincueq}
\ee
Here $\bra M^2 \ket_L$ and $\bra M^4 \ket_L$ denote the second and the fourth moments of the
probability distribution of the magnetization $P_L(M)$, where
\be
\bra M^k \ket_L=\int dM M^k P_L(M)
\label{moments}
\ee
In the ordered phase $V_L \to \frac{2}{3}$.
An appropriate method for determining the transition temperature $T_c$ is to record the variation of
$V_L$ with $T$ for various system sizes and then locate the intersection of these curves.
\begin{figure}[!h]
\centering
\resizebox{80mm}{!}{\includegraphics[scale=0.6]{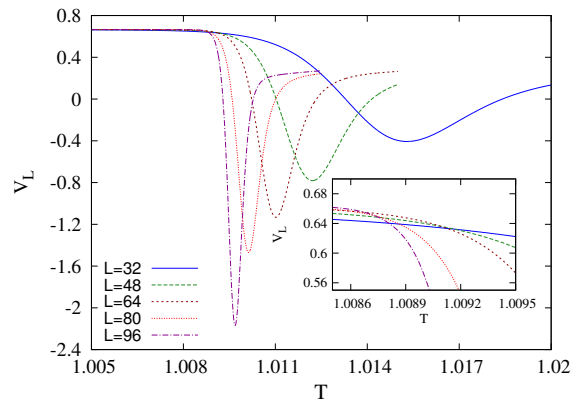}}
\caption{(Color online) Fourth-order cumulant of magnetization plotted against temperature
$T$ for different lattice sizes across a first order phase transition. The inset is the 
enlargement of $V_L$ around the transition temperature.}
\label{bincu}
\end{figure}
In Fig. \ref{bincu} we show the Binder's cumulant across a first order phase transition for
various lattice sizes. The inset of Fig. \ref{bincu} shows the same for a smaller range of
temperature.
One compares the values of $V_L$ for two different lattice sizes $L$ and $L^{\prime} = bL$, 
making use of the condition 
\be
\left(V_{bL}/V_L\right)_{T=T_c}=1
\label{bincucond}
\ee
Because of the presence of residual corrections due to finite size scaling, one actually
needs to extrapolate the results of this method for $(\ln b)^{-1} \to 0$ \cite{bin}. For each lattice
size we obtained the optimized distribution which was then used to calculate the 
cumulant $V_L (T)$ in the critical region. Due to corrections to scaling, the estimates for
the transition temperature $T_c$ depend on the scale factor $b=L^{\prime}/L$ so that the
extrapolation procedure is necessary. Results of the extrapolation are shown in 
Fig. \ref{bincutc}.
The thermodynamic limit of the transition temperature is found to be $1.00866 \pm 0.00013$.
For the plot of Fig. \ref{bincutc}, we have taken $L = 32$ and $L^{\prime} = 48$, $64$, 
$80$ and $96$ respectively. The transition temperatures estimated from the peak position
of the susceptibility and the Binder's cumulant method differs by only $0.04 \%$.

Recently it was shown \cite{ss4} that for strong enough non linearity ({\it i.e.}, for large values
of $p^2$) in the interaction potential of Eq. (\ref{pot}), there is a sudden proliferation of
topological defects that makes the system disordered and consequently the transition is 
associated with a discontinuous non universal jump in the helicity modulus. Thus the present work
supports the idea that the type of phase transition in thin superconducting films may be changed
due to the influence of disorder. It may be noted that the effect of disorder on the KT transition
has become relevant since the experimental observation of the superconductor-insulator transition
in thin disordered films \cite{liugold,paala}.
In this work we have explored how the reweighting of numerical data obtained in extensive
MC simulations with $16 \le L \le 96$, together with Lee-Kosterlitz's method of finite
size scaling and analysis of Binder's cumulant yield useful information about the equilibrium
critical properties of the classical XY model with a modified form of interaction potential.
The MHR method is extended to calculate quantities other than energy, specific heat etc without
constructing the two-dimensional histograms. This approach is economic in terms of computer memory
and CPU time as well, and is thus not trivial in scope. We have applied this method to calculate
magnetization, susceptibility and Binder's fourth order cumulant of magnetization to a system which 
is relatively harder to simulate because of the presence of a deep and narrow potential well. The 
method can be applied to any lattice spin system with discrete as well as continuous energy spectra.
Since there are no limitations on the method of the simulation, this approach could also be useful
for simulations in chemistry and biology. 

\section{Acknowledgements}
The author acknowledges support from the UGC Dr. D. S.
Kothari Post Doctoral Fellowship under grant No. F.4-2/2006(BSR)/13-416/2011(BSR). I
thank S. K. Roy for a critical reading of the manuscript.

\end{document}